# Push Puppet Networks

A Structured Bayesian Pruning Algorithm for Language Model Compression

Robert Kubinec

## Author Note

Robert Kubinec https://orcid.org/0000-0001-6655-4119.

Access the code, data, and analysis at https://github.com/saudiwin/push_puppet.

Correspondence concerning this article should be addressed to Robert Kubinec

(bobkubinec\@gmail.com)

**Abstract**

This paper presents push puppet networks, a novel Bayesian algorithm for structured pruning of large language models. The push puppet network learns a hierarchical function during training that can optimally determine specific network layers to keep for a given target size. By adding a small number of gating parameters via a hierarchical penalty function, the learned smooth function can allow for a network to be resized to very specific sizes without loading the full model into memory or requiring further post-training computation. The method compares favorably with existing approaches (SparseGPT, Wanda) at high pruning sizes (less than 50% of network structure) while realizing measurable speed-ups on conventional GPUs with PyTorch. Furthermore, push puppet networks can achieve significant speedups as candidates for speculative decoding.[1]

[1]I thank the National Science Foundation ACCESS program and Texas A&M's ACES cluster for computational resources for this project.

# Push Puppet Networks

## Conflicts of Interests

The author has no conflicts of interests to declare.

## AI Usage Statement

Claude Code with Opus 4.8 was used for writing code to implement the models and ensured that the mathematical formal exposition matched the code. The author wrote the paper while Opus 4.8 provided feedback on the draft and checked the equations for consistency.

Sparse computation for large language models (LLMs) has become an important research endeavor given the high costs of running LLMs at scale. This paper presents a novel method for structured pruning of LLMs; that is, for compression that is realistic under contemporary hardware, in particular graphical processing units (GPUs) that work best with regular-sized matrices. This algorithm borrows its design from the idea of a push puppet, which is a child's doll that can disassemble and reassemble itself in response to the varying pressure of a string. The string in this context is a smooth learned hyper-function (i.e., a Bayesian prior) that contains information as to which neurons contain the most information for a given sparsity level. This function permits structured pruning from very minor sparsity to almost complete sparsity (up to only 5% of model layers remaining).

Relative to existing methods like sparseGPT (Frantar and Alistarh 2023) and Wanda (Sun et al. 2024), this method implements structured pruning, and such, it does not require custom kernels for sparse computation beyond what is available in platforms like PyTorch. Furthermore, as I show, this method is much more robust to degradation at high levels of sparsity (i.e., below 40%), which makes it an ideal method for designing target models for speculative decoding. Furthermore, because the pruning method is known *before* the model is loaded into memory, the method avoids memory bottle necks associated with sparseGPT and Wanda that require the model to be loaded in memory before pruning.

In this paper, I compare the algorithm's performance on training data from the Olmo-3 project (Olmo et al. 2026) for a model of 1.3b parameters. Furthermore, I show how this algorithm permits much more flexible speculative decoding, which enables users to employ speculative decoding on consumer-grade hardware (Leviathan et al. 2023).

## Algorithm Definition

This method relies on Bayesian concepts, especially that of an informative prior (Gelman et al. 2013), to identify a more flexible form of sparse computation. While many pruning methods are done after a model is trained in order to prune it to a certain size (Lee et al. 2026), such as Wanda (Sun et al. 2024), sparseGPT (Frantar and Alistarh 2023), and model distillation (Lee et al. 2026), this method seeks to learn a *continuous* function of optimal sparsity. To do so, I employ a parameter $\lambda$ that represents the cost of computation and vary this parameter during the training period according to the following distribution:

$$\lambda \sim \text{Exponential}\ (0.3) \tag{1}$$

For each step of the stochastic optimization procedure, a new value of $\lambda$ is drawn so that the model can adjust to the full range of possibilities from very small values of $\lambda$ to relatively large values. Because the draws of $\lambda$ are stochastic over batches of data in gradient descent, the model must learn to be robust to any plausible value of $\lambda$ given that the penalty function is jointly estimated with the model parameters.

The specific penalty function for $\lambda$ as the cost of computation will vary according to the particular model architecture. In the mathematical formulation that follows, I assume a deep neural network of reasonable complexity with a transformer architecture and an attention layer that follows Olmo et al. (2026). Specifically, we can begin with an autoregressive transformer with $L$ blocks, vocabulary size $V$, model dimension $d$, $H$ attention heads, and FFN intermediate dimension $d_{\text{ff}}$.

Given an input token sequence $\mathbf{u} = (u_1, ..., u_T)$, we start with an embedding layer defined as:

$$\mathbf{x}_0 = \mathbf{E}[\mathbf{u}] + \mathbf{P} \tag{2}$$

where $\mathbf{E} \in \mathbb{R}^{V \times d}$ is the token embedding matrix and $\mathbf{P} \in \mathbb{R}^{T \times d}$ is the learned positional embedding.

The FFN layer uses a SwiGLU activation with three projection matrices:

$$\text{FFN }(\mathbf{x}) = \mathbf{W}^{\text{down}}(\text{SiLU }(\mathbf{W}^{\text{gate}}\mathbf{x}) \odot \mathbf{W}^{\text{up}}\mathbf{x}) \tag{3}$$

with $\mathbf{W}^{\text{gate}}, \mathbf{W}^{\text{up}} \in \mathbb{R}^{d_{\text{ff}} \times d}$ and $\mathbf{W}^{\text{down}} \in \mathbb{R}^{d \times d_{\text{ff}}}$.

$Q$, $K$, $V$ are projected separately from the input and split across $H$ heads with head dimension $d_h = d/H$. Per-head RMSNorm is applied to $Q$ and $K$ after projection:

$$\mathbf{Q} = \text{Norm}_h\big(\mathbf{x}\,(\mathbf{W}^Q)^\top\big), \quad \mathbf{K} = \text{Norm}_h\big(\mathbf{x}\,(\mathbf{W}^K)^\top\big), \quad \mathbf{V} = \mathbf{x}\,(\mathbf{W}^V)^\top, \tag{4}$$

$$\mathbf{W}^Q, \mathbf{W}^K, \mathbf{W}^V \in \mathbb{R}^{d \times d} \tag{5}$$

where $\text{Norm}_h$ is RMSNorm applied independently to each head's $d_h$-dimensional slice.

$$\text{Attn }(\mathbf{x}) = \text{concat}_{h=1}^{H}\left[g(\left(\frac{\mathbf{Q}_h \mathbf{K}_h^\top}{\sqrt{d_h}} + \mathbf{M}_{\text{causal}}\right)\mathbf{V}_h)\right]\mathbf{W}^O, \quad \mathbf{W}^O \in \mathbb{R}^{d \times d} \tag{6}$$

where $\mathbf{M}_{\text{causal}}$ is the upper-triangular $-\infty$ mask stored as a fixed buffer and $g(\cdot)$ is the softmax function.

The intention of this model definition is to represent a relatively generic formulation that matches recent large language models in production. To incorporate our computational cost parameter $\lambda$, we add a separate penalty function to the calculation of cross-entropy loss $NLL$:

$$\mathcal{L} = \mathrm{NLL}_{\mathrm{eff}} + \lambda \cdot \bar{\rho}(\lambda) \tag{7}$$

where $\mathrm{NLL}_{\mathrm{eff}}$ is the scaled cross-entropy loss and $\bar{\rho}(\lambda)$ is the weighted mean inclusion probability of the attention and FFN layers. Importantly, $\lambda$ sets the level of the penalty exogenously as a draw from the exponential distribution *and* affects the probability of parameter inclusion $\bar{\rho}(\lambda)$. This multiple influence of $\lambda$ on the training objective is what permits the algorithm to adjust to very high and very low penalties effectively.

In order to make the penalty equal a computational cost, the penalty is weighted by the relative computational burden of different components of the model. Specifically, the normalized penalty is:

$$\bar{\rho} = \frac{\sum_\ell w_{\mathrm{ffn}} \bar{\rho}_\ell^{\mathrm{ffn}} + \sum_\ell w_{\mathrm{attn}} \bar{\rho}_\ell^{\mathrm{attn}}}{\sum_\ell w_{\mathrm{ffn}} + \sum_\ell w_{\mathrm{attn}}} \tag{8}$$

where the FFN weight $w_{\mathrm{ffn}}$ is $6 \cdot d \cdot d_{\mathrm{ff}}$ and the attention weight $w_{\mathrm{attn}}$ is $8 \cdot d^2$. The factor of 6 reflects the three SwiGLU projection matrices (gate, up, and down), each contributing $2 \cdot d \cdot d_{\mathrm{ff}}$ multiply-accumulate operations; $w_{\mathrm{attn}} = 8d^2$ reflects the four attention projections ($Q$, $K$, $V$, and output).

In addition to these weights, what drives the penalty is the inclusion probability $\rho(\cdot)$ for a given row/layer of the FFN and attention components where the weight matrix for each type has one row per intermediate neuron.

The $\ell_2$ norm of each row measures how much that neuron can contribute to the output:

$$r_i = \| \mathbf{w}_i \|_2, \quad i = 1, ..., d_{ff} \tag{9}$$

so that important neurons receive a higher inclusion probability via the inclusion function $\rho_i(\lambda)$:

$$\rho_i(\lambda) = \sigma\left(\frac{r_i - \tau(\lambda)}{T}\right) \tag{10}$$

where $\sigma$ is the logistic function and $T > 0$ is a fixed temperature controlling the sharpness of the inclusion threshold (in the implementation $T = 0.5$).

If $r_i$ is high, then the inclusion probability for that row increases. The inclusion function is an analogue of the item-response theory parameterization (Kubinec 2026) except that $\lambda$ is set exogenously rather than learned. Importantly, *either* the row can be important or the the $\tau(\cdot)$ function can be low to allow a given row in the weight matrix to be included in the model. Consequently, there are two distinct mechanisms through which the model can adjust to computational costs. This learned flexibility is crucial to allow the network to self-prune while minimizing loss in predictive performance.

### The $\tau$ function

The $\tau(\lambda)$ is a crucial part of this model as it represents the partially learned, partially exogenous barrier to parameter inclusion. As such, this function must be carefully parameterized to avoid corner solutions, such as very low values where all parameters are included or none are included. Furthermore, the function must have a monotonicity restriction to prevent the function from paradoxically shrinking and re-growing the network in response to increasing $\lambda$, which does not fit with the idea of $\lambda$ as a computational cost parameter.

As such, a monotone third-order B-spline formulation is imposed on $\tau(\lambda)$ to allow for non-linear flexibility while also requiring smoothness. In other words, higher $\lambda$ values need to reduce the inclusion probabilities of rows in the weight matrices even if there may be points in the function that are slowly increasing or nearly flat. To create a monotone B-spline, the coefficients need to be transformed so that each coefficient is strictly greater than the previous one (i.e., the softplus transformation).

Each layer in the network has a separate B-spline function $\tau^{(\ell)}(\lambda)$ with $n_{\text{knots}} = 10$ control points, of which $n_{\text{knots}} - 1 = 9$ are learnable. The first control point is fixed to $-1/T$ (using the same temperature as Equation 10); at $\lambda = 0$ this places the threshold well below any typical row norm, ensuring all neurons are included. The remaining 9 coefficients are built by cumulative sum of softplus-transformed increments:

$$c_0 = -\frac{1}{T} \quad \text{(fixed)}, \qquad c_k = c_{k-1} + \text{softplus}\;(\theta_k), \quad k = 1, ..., 9 \tag{11}$$

The threshold for each neural head is the B-spline evaluated at these 10 control points:

$$\tau^{(\ell)}(\lambda) = B^{(\ell)}\left(\frac{\lambda}{\lambda+1}\right) \tag{12}$$

As will be shown later, this function is non-linear but still tractable and adds relatively little computational overhead. With the B-spline function, the model learns the optimal inclusion probability for each row $i$ in the FFN and attention weight matrices that is a function of *both* the row importance and the $\lambda$-influenced $\tau^{(\ell)}(\lambda)$ value. By so doing, the model can better avoid performance degradation as $\lambda$ increases by ensuring the most important neurons are included in the model.

## Amplitude and Input Scaling

The final step of the $\lambda$ penalty is to include a scaling function that can adjust the amplitude and direction of the weights in the model depending on their probability of inclusion. This step helps the model compensate for some neurons being dropped from computation, and provides an additional mechanism through which the model can adjust to varying computational costs.

For FFN layers, each layer learns three independent amplitude corrections. Each uses its own monotone B-spline $u^{(\cdot)}(\lambda)$ initialized to approximate the rational map $\lambda/(\lambda+1)$ and trained jointly with all other parameters:

$$f_i^{\text{gate}}(\lambda) = e^{s_i^{\text{gate}} \cdot u^{\text{gate}}(\lambda)}, \quad f_i^{\text{scale}}(\lambda) = e^{s_i^{\text{out}} \cdot u^{\text{scale}}(\lambda)}, \quad f_j^{\text{input}}(\lambda) = e^{s_j^{\text{in}} \cdot u^{\text{input}}(\lambda)} \tag{13}$$

where $s^{\text{gate}}, s^{\text{out}} \in \mathbb{R}^{d_{\text{ff}}}$ are per-neuron scalars and $s^{\text{in}} \in \mathbb{R}^{d}$ is a per-input-dimension scalar.

The effective weight of the gate matrix at inference uses the output scale $f^{\text{scale}}$ and input scale $f^{\text{input}}$:

$$w_{ij}^{\text{eff}}(\lambda) = w_{ij} \cdot m_i \cdot f_i^{\text{scale}}(\lambda) \cdot f_j^{\text{input}}(\lambda) \tag{14}$$

Because $f^{\text{input}}$ varies across input dimensions the correction is a separable (rank-1) matrix $f^{\text{scale}}(f^{\text{input}})^\top$. This allows the model to modulate both the magnitude and direction of surviving neurons, which is similar in spirit to the Hessian correction in SparseGPT.

The inclusion probability uses the gate scale $f^{\text{gate}}$, which is a separate learned parameter from $f^{\text{scale}}$:

$$\rho_i(\lambda) = \sigma\left(\frac{f_i^{\text{gate}}(\lambda) \cdot \| w_i^{\text{gate}} \odot f^{\text{input}}(\lambda) \|_2 - \tau^{(\ell)}(\lambda)}{T}\right) \tag{15}$$

where $w_i^{\text{gate}}$ is the $i$-th row of $\mathbf{W}^{\text{gate}}$. The extra parameters are $2d_{\text{ff}} + d$ scalars per FFN layer plus three B-spline parameter vectors ($n_{\text{knots}}$ values each), which represents relatively few parameters compared to a deep network of any appreciable size.

Attention heads use the same two-role split as FFN layers but without input-direction scaling:

$$f_h^{\text{gate}}(\lambda) = e^{s_h^{\text{gate}} \cdot u^{\text{gate}}(\lambda)}, \quad f_h^{\text{scale}}(\lambda) = e^{s_h^{\text{out}} \cdot u^{\text{scale}}(\lambda)} \tag{16}$$

with inclusion probability $\rho_h(\lambda) = \sigma\left(\frac{\|\text{head}_h\|_F \cdot f_h^{\text{gate}}(\lambda) - \tau^{(\ell)}(\lambda)}{T}\right)$, and effective head outputs scaled by $f_h^{\text{scale}}(\lambda)$.

## Training

Training optimizes the model jointly with the $\lambda$ penalty functions by minimizing the combined loss (Equation 7). In order to train the model to drop neurons, at each forward pass a 0/1 mask $m_i$ is drawn from a Bernoulli distribution where $p = \rho_i$ as defined in Equation 10 given the sampled value of $\lambda$, that is, the estimated inclusion probability:

$$m_i \sim \text{Bernoulli}\ (\rho_i), \quad m_i \in \{0, 1\} \tag{17}$$

A straight-through estimator (STE) is used to create a continuous gradient given the hard mask:

$$\tilde{m}_i = m_i + (\rho_i - \rho_i^{\text{sg}}) \tag{18}$$

where $(\cdot)^{\text{sg}}$ denotes stop-gradient. This creates a hard effective mask $\tilde{m}_i = m_i$ for the forward pass and a smooth $\partial \tilde{m}_i / \partial \rho_i = 1$ gradient in the backward pass. The cross-entropy $\text{NLL}_{\text{eff}}$ is the standard token-level cross-entropy loss computed over all tokens regardless of the mask, which encourages the model to balance the $\lambda$ penalty and predictive performance. Surviving neurons have their output amplitude adjusted by $f_i^{\text{scale}}(\lambda)$.

In order to train the model for full performance, $\frac{1}{3}$ of training steps are given $\lambda = 0$ while the other $\frac{2}{3}$ of steps receive $\lambda$ from the exponential distribution at the specified rate. This process ensures that the model is able to continuously prune from maximal to minimal performance conditional on $\lambda$.

Otherwise, training proceeds with an AdamW optimizer ($\beta_1 = 0.9$, $\beta_2 = 0.95$, weight decay 0.1) with cosine learning rate decay.

## Inference

To enable structured pruning to a specific size, the user must first choose a $\lambda$ value, which is then passed to the B-spline functions (both for inclusion probability and amplitude correction). Once the inclusion probability is calculated, a sample of weight rows $r_i$ is drawn ($m_i$), and then the model is structurally pruned so that only surviving rows $r_i$ are stored in memory for both FFN layers and attention heads. This can be done before a model is loaded into memory to avoid hitting memory bottlenecks in either GPUs or CPUs.

In order to determine the size of a model conditional on $\lambda$, the fitted model must be pruned at varying sizes of $\lambda$ to determine the expected surviving parameter counts. However, once that relationship is estimated, it is trivial to select a $\lambda$ value to obtain a sample of model weights that is equal to the expected size.

At higher values of $\lambda$ and small expected parameter counts, prediction error for the model will increase significantly and the stochastic nature of mask sampling $m_i$ will also increase. In that case, it is possible to sample multiple masks $m_i$ and average the model predictions to reduce uncertainty related to mask sampling (though multiple masks cannot be used to approach dense model performance).

Evaluation

To evaluate the algorithm, I fit the previously described model to a sample of approximately 5 billion tokens from the Olmo-3 open dataset (Olmo et al. 2026). All models use the OLMo-3 transformer architecture with $d = 2048$, $H = 16$ attention heads, $L = 16$ layers, an FFN intermediate dimension of $d_{\text{ff}} = 8192$, and a vocabulary of 100,278 tokens (approximately 1.3 billion total parameters), trained at a sequence length of 1,024. The model is trained for 3 epochs with $\lambda \sim \text{Exp}\ (0.3)$, sampling $\lambda = 0$ (dense pass) with probability $\frac{1}{3}$ and $\lambda > 0$ with probability $\frac{2}{3}$. This model size is chosen so that size constraints matter but estimation is still within computational tractability. Given the model's relatively small size and limited training regime, results will focus on perplexity of token prediction rather than task-related benchmarks.

## Lambda Response Curve

We first examine how model performance changes for varying $\lambda$ values by pruning the trained model in terms of attention and FFN layers for a given $\lambda$. Figure 1 shows the change in perplexity (PPL) as $\lambda$ increases from zero (dense, no pruning) to 5.0, which is near the upper end of the learned funciton. Each point is the mean perplexity averaged from 5 independent mask draws at a given $\lambda$ value; the dotted line shows the dense model performance as a benchmark.

The plot shows that there is a non-linear yet reasonably smooth degradation of model performance as FFN layers and attention heads are dropped. The smoothness is driven by the B-spline function definition and the computational penalty imposed on the model's loss function.

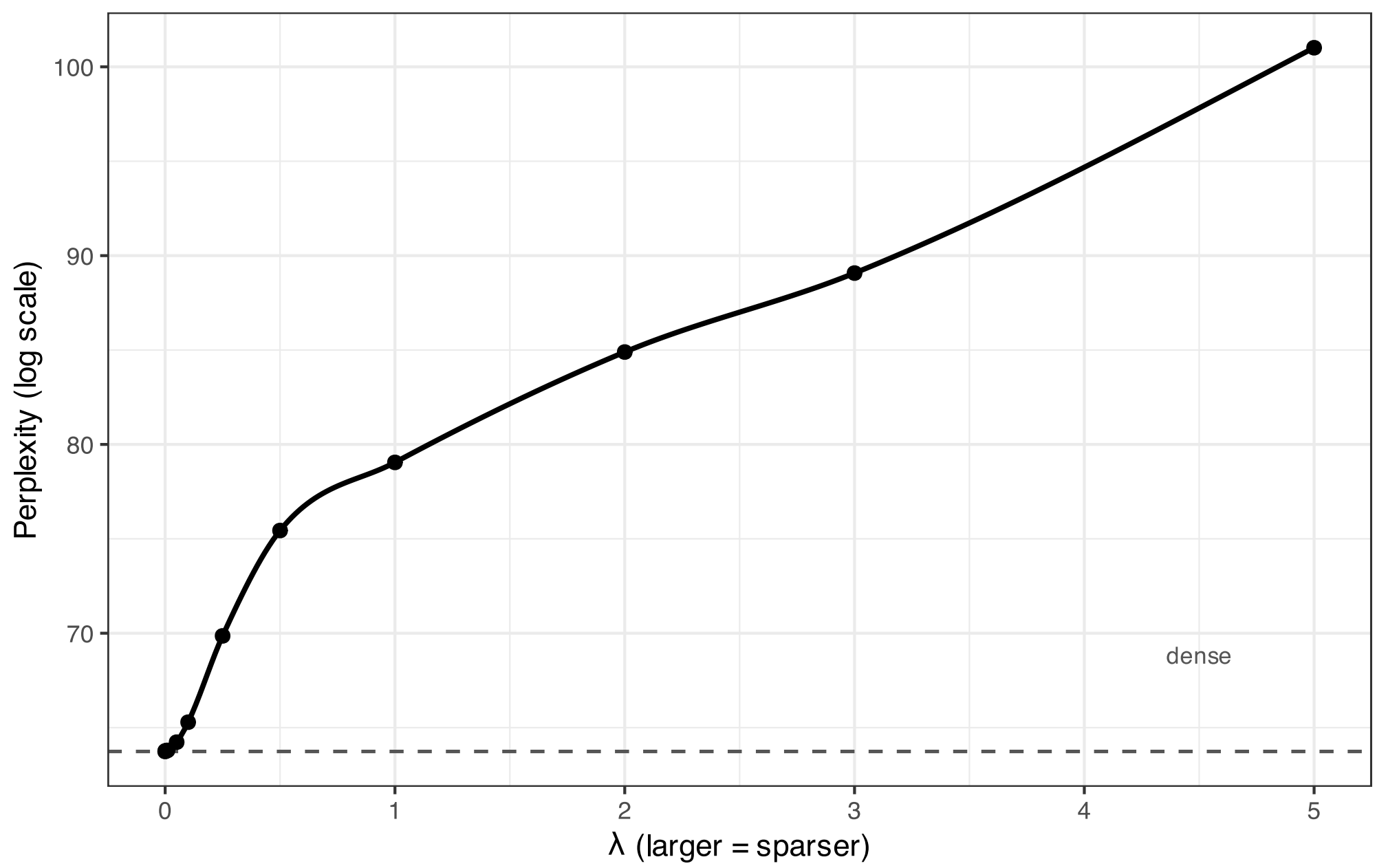

Figure 1: Perplexity as a function of a given value of the sparsity budget $\lambda$. The dashed line marks the dense baseline. Points show mean perplexity over 10 mask draws at each $\lambda$; lines are monotone spline interpolants.

## Active Unit Fractions

To make it clear how the $\lambda$ function drives structured pruning, Figure 2 shows the proportion of FFN neurons (solid lines) and attention heads (dashed lines) that survive pruning at each $\lambda$ level. These parameters only survive if they are selected by the mask $m_i$ for a given $\lambda$.

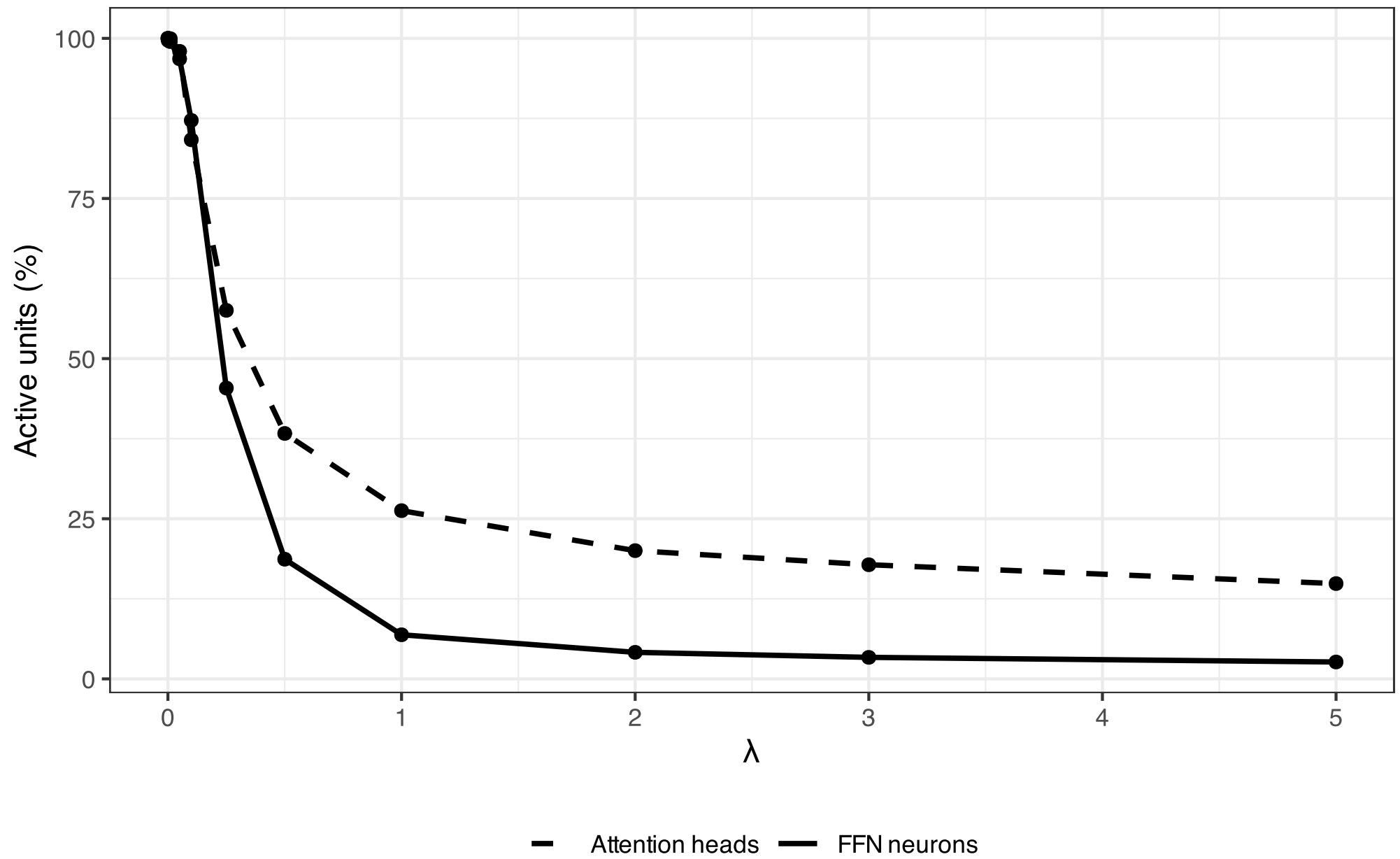

Figure 2: Solid lines show the fraction of FFN neurons retained; dashed lines show the fraction of attention heads retained. The FLOP-weighted penalty causes attention heads to be pruned more conservatively than FFN neurons at equivalent $\lambda$ values.

## Sampling Error

The stochastic nature of the mask $m_i$ means that as the model becomes more sparse, there is residual error in terms of which layers are kept in the model. Figure 3 shows the mean and range of perplexity across 30 independent mask draws at the average value of $\lambda$ from the training runs (approximately 3). While this error does exist, it is within reasonable bounds even at the relatively high value of $\lambda = 3$.

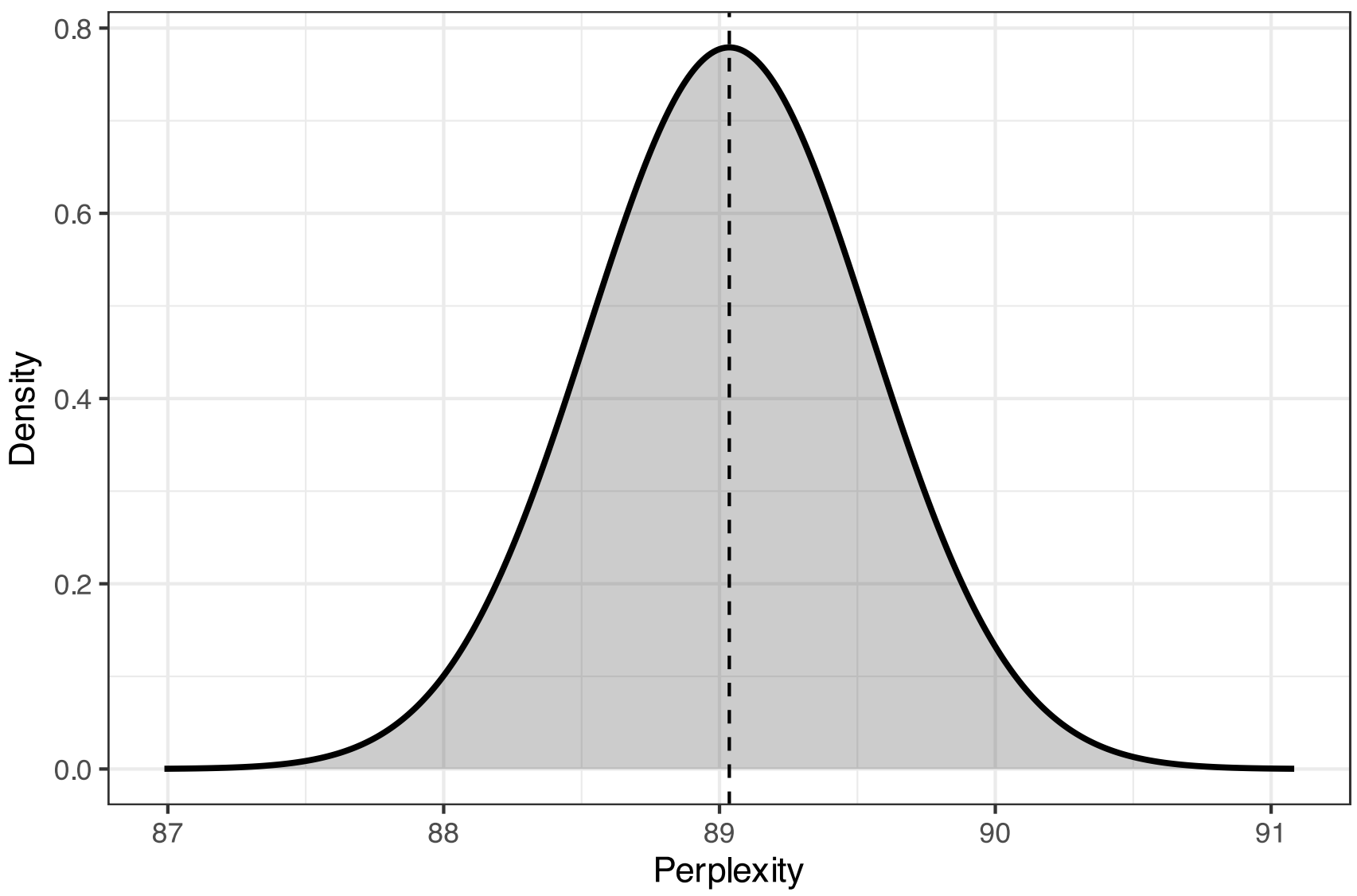

Figure 3: Distribution of perplexity across mask draws at the training-mean $\lambda \approx 3$, modelled as a Normal density $\mathcal{N}(\text{ppl_mean}, \text{ppl_std})$ whose mean and standard deviation are estimated from 30 independent resamples.

## Learned Threshold Behaviour

Figure 4 shows how the learned threshold $\tau(\lambda)$ changes in response to $\lambda$, revealing a similar non-linear yet smooth pattern. Importantly, this plot does not show any performance changes but rather shows the internal model relationship between the penalty $\lambda$ and the inclusion factor $\tau$, which ultimately determines how many neurons survive (given their importance).

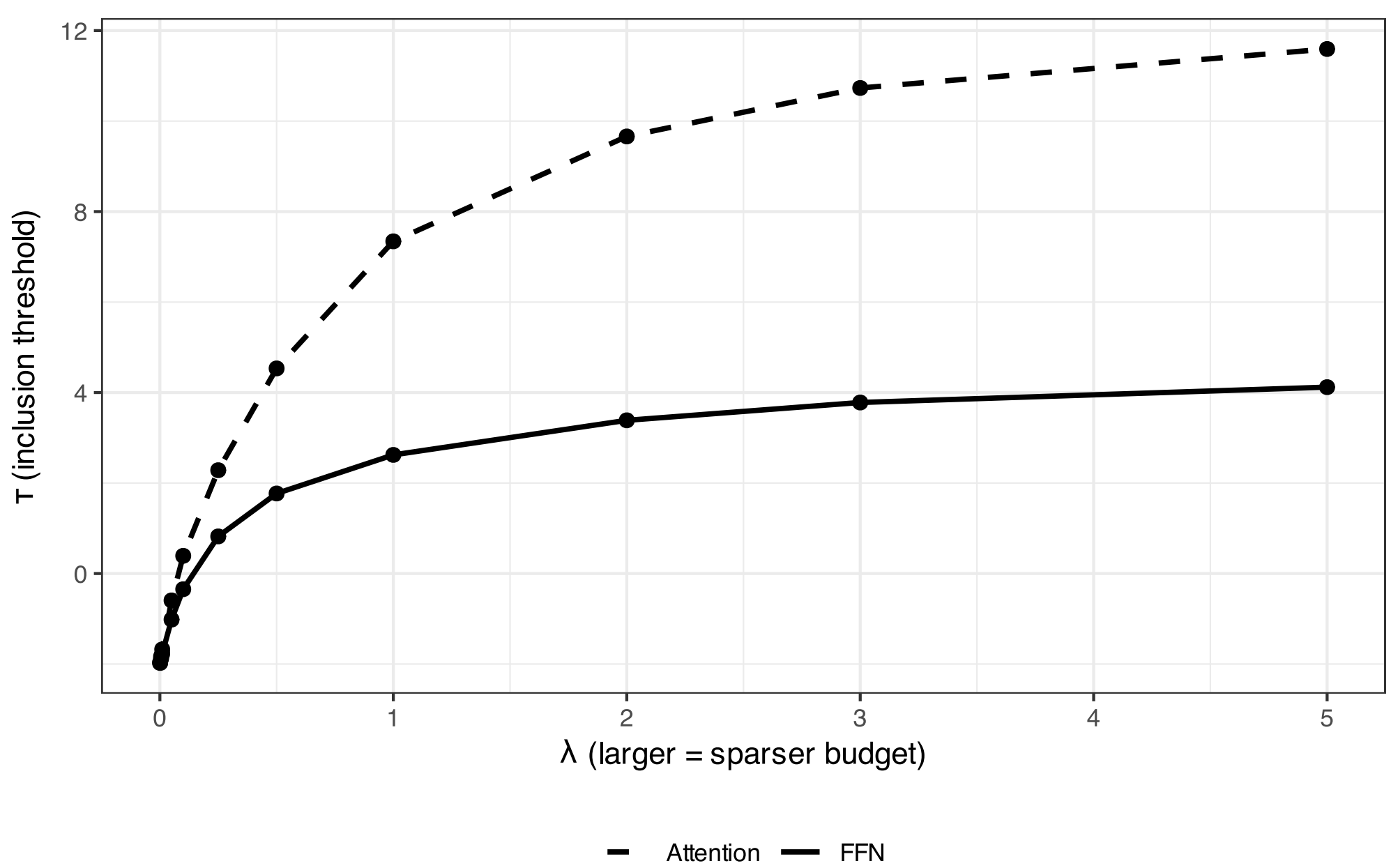

Figure 4: Learned inclusion threshold $\tau$ vs $\lambda$.

## Compression–Quality Frontier

Finally, Figure 5 shows the trade-off between model size (in MB after pruning either FFN layers or attention heads) and PPL. Each line traces the size and PPL performance of the two model variants given $\lambda$ values. As can be seen, the model can be compressed from 5 GB to approximately 1.3 GB before PPL performance decreases dramatically, showing the power of the algorithm at identifying reasonably performant sub-networks at very high sparsity.

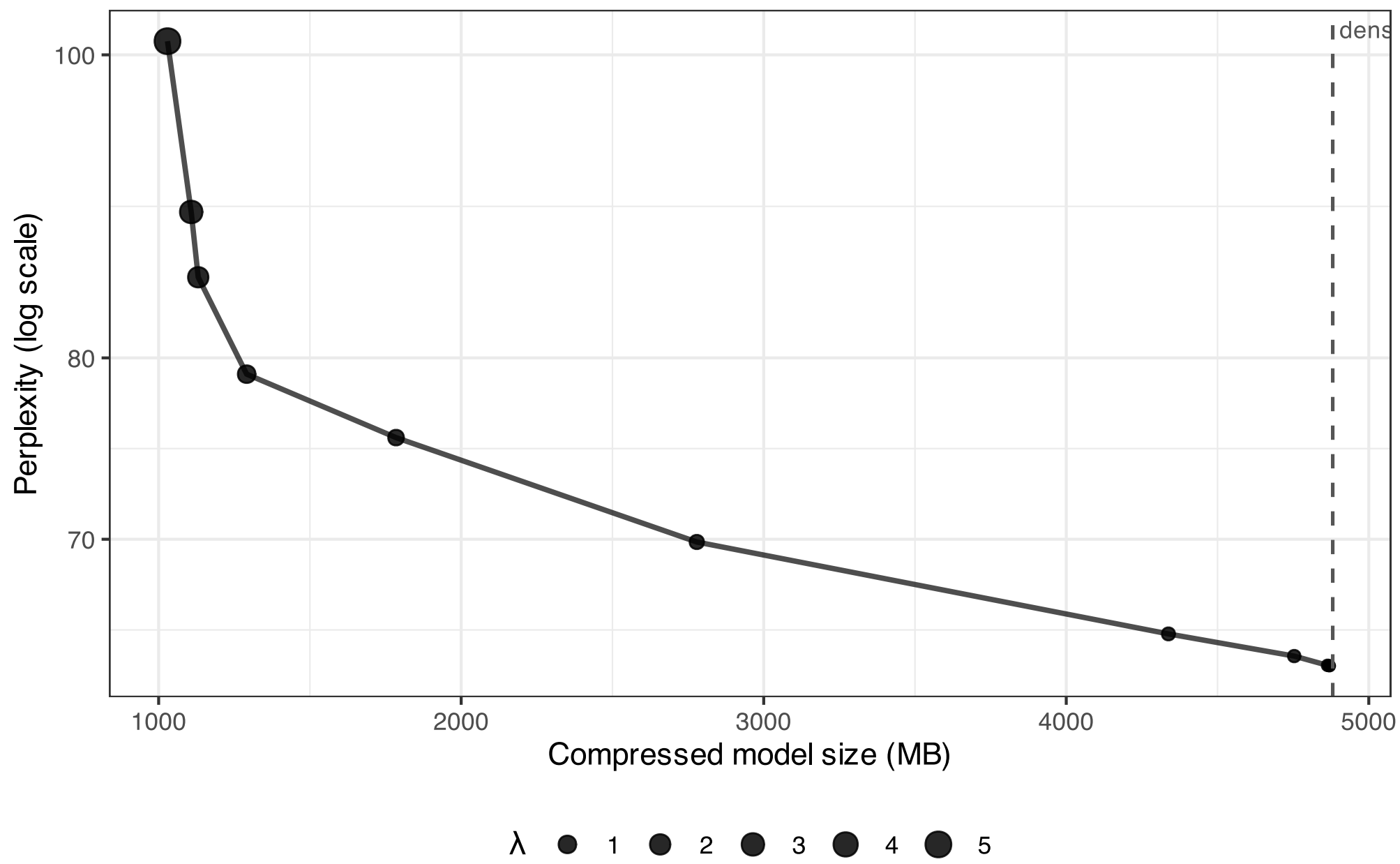

Figure 5: **Compression–quality frontier.** Perplexity vs compressed model size (MB) after physical removal of pruned rows and columns. The curve traces `hyper_joint` across $\lambda$ values; point size scales with $\lambda$. The vertical dashed line marks the dense model size. Lower-left is better.

## Inference Speedup

In order to make the compression figures concrete, Figure 6 calculates the actual evaluation speedup of compressing the push puppet network on an Intel Data Center Max GPU (48 GB RAM). As can be seen, for this model the computational speed-ups max out at about 4x around a $\lambda$ value of 1.5. The speed-ups do not increase above this value mainly because the model has become so small that additional speed-ups are not computationally noticeable.

Nonetheless, it is important to note that these speedups were actualized on realistic hardware. In order to handle the irregularly-sized weight matrices post-pruning, the `pytorch torch.compile` function was used to create optimal matrix multiplication code. While a custom kernel might show even better performance, we note that a 4x speedup on a small model is a significant improvement, and at $\lambda = 1.5$ the model has increased in PPL by only 30%.

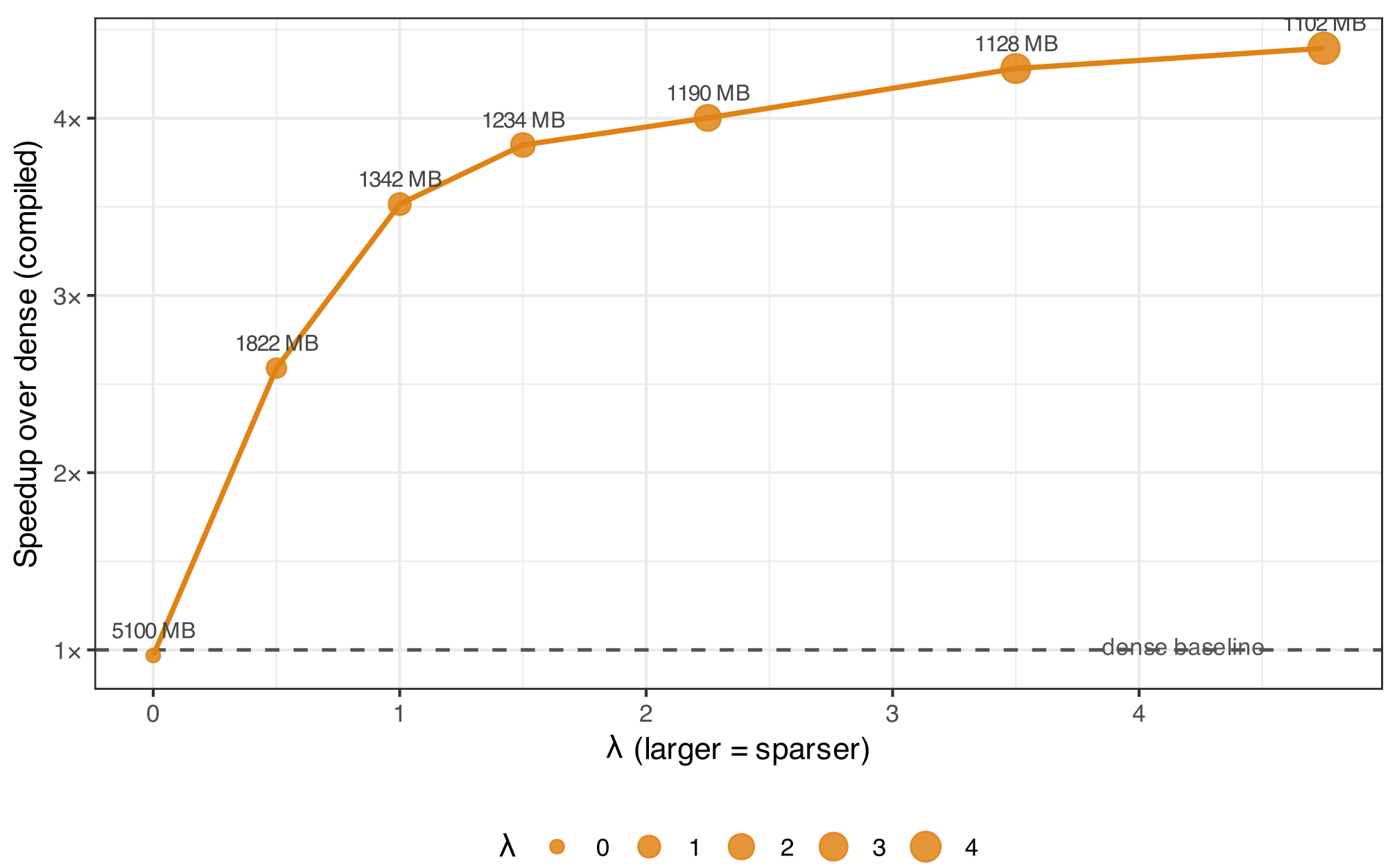

Figure 6: Speedup of the compiled push puppet sub-network versus the compiled dense baseline (dashed line at 1×) at each evaluated $\lambda$. Points are sized by $\lambda$. The right-hand axis shows the physically compressed model size in MB.

### Comparison with Wanda and SparseGPT

In this section, I compare the $\lambda$-based sub-network sampling against SparseGPT (Frantar and Alistarh 2023) and Wanda (Sun et al. 2024) in terms of compression and model performance relative to the dense model, i.e., $\lambda = 0$. Wanda and SparseGPT are evaluated at 20 unstructured sparsity levels from 5% to 99% applied to the dense model with $\lambda = 0$ (i.e. fully dense); the model is then swept over $\lambda$ values from 0 to 5. Figure 7 and Figure 8 show the performance of these different methods in terms of the PPL cost over baseline and the relative memory and computation savings relative to each other and the dense model.

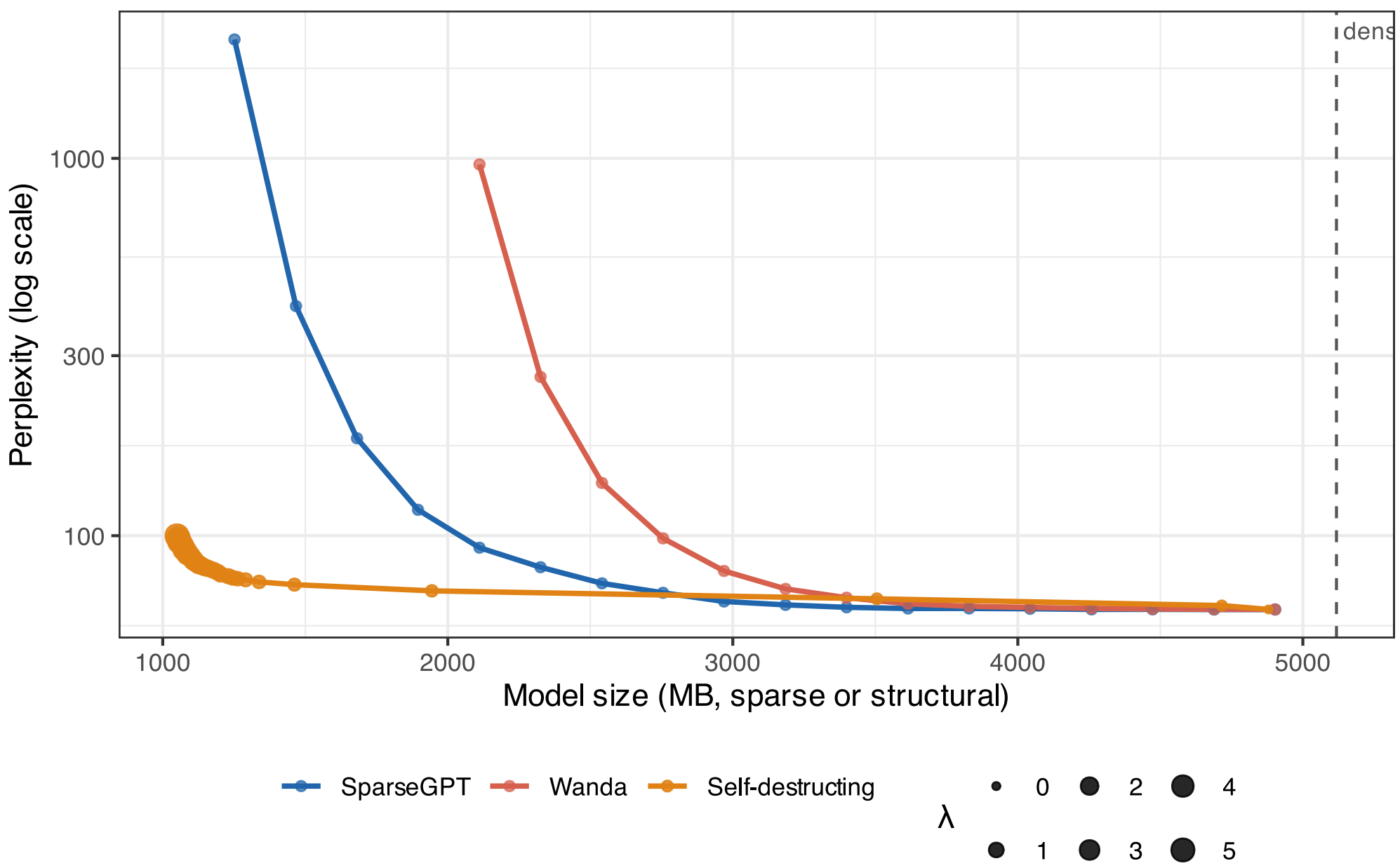

Figure 7: Each curve traces one method across its compression range. For SparseGPT and Wanda, x-axis is the non-zero-weight size at sparsity $s$ (i.e., $(1 - s) \times$ dense MB assuming sparse storage). For the push puppet method, x-axis is the physically compressed size after structural row/column removal. Points with PPL > 3,000 (model collapse) are excluded. Lower-left is better.

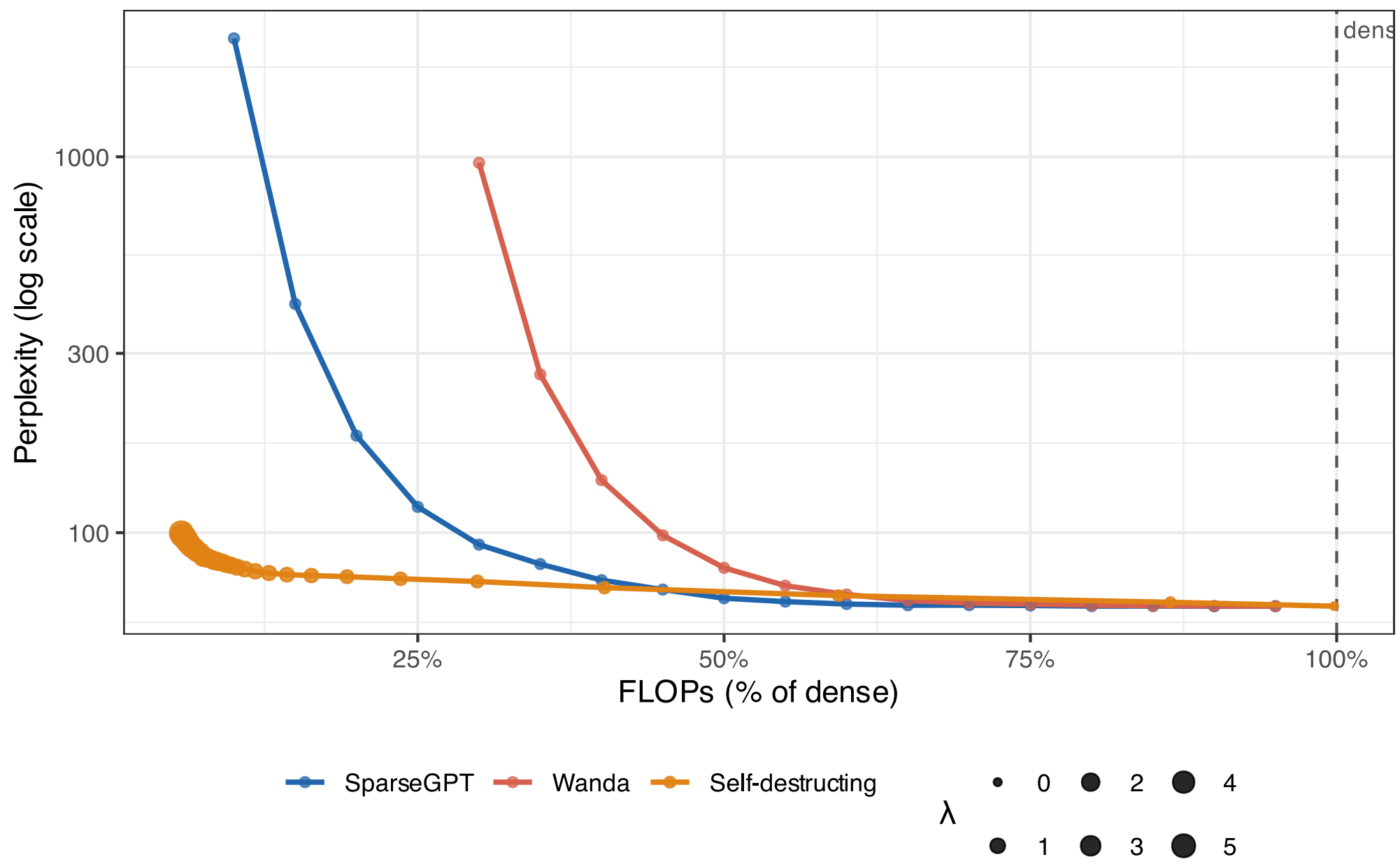

Figure 8: For SparseGPT and Wanda, FLOPs fraction equals $(1 - s)$ at unstructured sparsity $s$, assuming ideal sparse execution. For the push puppet method, FLOPs are computed from the FLOP-weighted mean of surviving FFN neurons and attention heads.

As can be seen, Wanda and SparseGPT out-perform the push puppet method at low levels of sparsity, while the push puppet outperforms at higher levels of sparsity. While the threshold will depend on the particular model, it would seem that the push puppet method dominates above 50% sparsity while the Wanda and SparseGPT have better performance at lower levels of sparsity (i.e., relatively dense models).

However, it is important to note that SparseGPT and Wanda are unstructured pruning methods; i.e., they will zero out particular neurons in the network regardless of where they are situated. While this approach is very granular, it does not result in actual computation savings in many cases because GPU kernels cannot handle sparse matrices. A notable exception is Nvidia's support for 2:4 sparsity levels in which 2 out of every 4 coefficients are zeroed out; otherwise, these methods' sparsity savings are theoretical.

By contrast, the push puppet $\lambda$ function uses structured pruning of specific rows in weight matrices (as shown in the previous section) so that the savings in terms of memory size and computation can be utilized without using custom kernels. Furthermore, the push puppet method can be implemented without loading a model into memory by sampling neurons at a given $\lambda$ level and then loading those specific weight matrices; sparseGPT and Wanda require the model to be loaded in-memory in order to calculate pruning. In addition, it is important to note that the push puppet method can be fruitfully combined with sparseGPT in which a push puppet model can be first trimmed with sparseGPT to the 2:4 sparsity ratio and then further pruned using $\lambda$ to achieve even greater sparsity.

**Speculative Decoding**

The push puppet method is particularly suited for making candidate models for speculative decoding (Leviathan et al. 2023) because push puppet sampled sub-networks are derived from the same model architecture as the dense (acceptance) model used for decoding. For this reason, the push puppet method can maintain high acceptance rates even at high structured sparsity levels, which permits good speculative speed-ups.

I first measure speculative performance by sampling sub-networks for varying $\lambda$ and using the dense model $\lambda = 0$ as the acceptance model. Figure 9 reports the *measured* end-to-end decoding speedup on an Intel Data Center Max (PVC) GPU with $\gamma = 4$ draft tokens per round. Each point is the mean generation throughput over three trials relative to the dense baseline; points are coloured by the measured token acceptance rate. As $\lambda_{\text{draft}}$ rises the draft becomes cheaper but performance degrades, so the acceptance rate falls with increasing sparsity. For these reasons, the speedup decreases as $\lambda$ increases and acceptance rates for proposals fall, though the speedup remains at around 2x even for only 20% of parameters remaining in the model. Furthermore, while the speed-up is 3x at very lower sparsity, it should be noted that this essentially amounts to running the model in parallel because the GPU has more than enough RAM to hold two full copies of the model. For these reasons, the 2x speed-up at high sparsity of 20% is the more interesting and potentially useful speed-up that could translate into production scenarios.

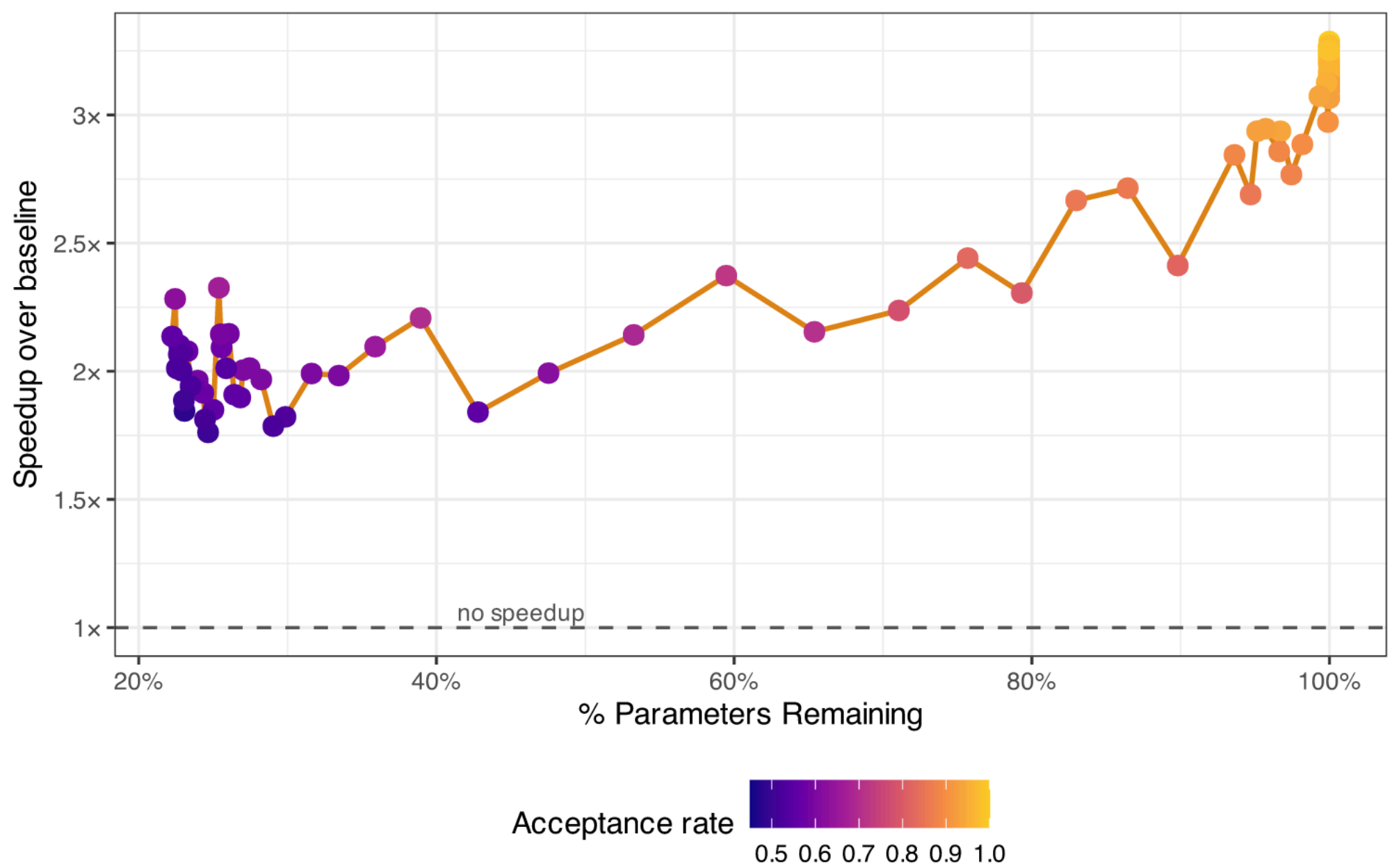

Figure 9: Measured decoding speedup of the push puppet sub-network with the dense ($\lambda = 0$) target, on an Intel Data Center Max GPU with $\gamma = 4$ draft tokens per round. Speedup is generation throughput relative to the naive autoregressive baseline (dashed line at 1 $\times$).

## Discussion

This paper presents the push puppet network method for structural pruning of large language models. It is evaluated based on a deep neural network architecture using Olmo et al. (2026), and thus is comparable to models used currently in production. At the same time, the method as presented must be incorporated during model training, and thus the results are shown for a model of smaller size and without many of the intensive post-training enhancements such as reinforcement learning.

It is certainly possible to consider learning a $\lambda - \tau$ function during a post-training re-estimation phase, but that is beyond the scope of this paper. What is clear is that the Bayesian hyper-function can learn how to decompose a network into remarkably stable sub-networks, and that these sub-networks can be used for sparse computation and for speculative decoding.